# Effects of RF Stimulus and Negative Feedback on Nonlinear Circuits

Renato Mariz de Moraes, and Steven M. Anlage

*Abstract*—We investigate the combined effect of rectification and nonlinear dynamics on the behavior of several simple nonlinear circuits. We consider the classic Resistor-Inductor-Diode (RLD) circuit driven by a low frequency source when an operational amplifier with negative feedback is added to the circuit. Radio frequency signals are applied to the circuit, causing significant changes in the onset of period-doubling and chaos. Measurements indicate that this effect is associated with a DC voltage induced by rectification of the RF signal in the circuit. The combination of rectification and nonlinear circuit dynamics produce qualitatively new behavior, which opens up a new channel of RF interference in circuits.

## I. Introduction

MOST work on electromagnetic interference (EMI) of circuits containing nonlinear elements based on the p-n junction, such as diodes and transistors, has focused on the effect of rectification [1]-[3]. The rectifying nature of the p-n junction causes the envelope of a high frequency signal to be stripped off and imposed on the circuit, sometimes with detrimental results. However, nonlinear dynamics, associated for example with the variable capacitance in the p-n junction, is present in addition to rectification. Hence rectification of an envelope signal alone [2] is not the whole story, and nonlinear dynamics of the circuit can introduce important new factors that can enrich the circuit analysis. It remains an open question whether nonlinear dynamics can create qualitatively new circuit behavior in the presence of external stimulus.

The driven Resistor-Inductor-Diode (RLD) circuit has been widely investigated because it is the simplest passive nonlinear circuit that displays period-doubling and chaos [4]-[9]. Although there is controversy over the precise mechanism of period-doubling in this circuit [6],[8],[10], it has become the standard representation of the nonlinear behavior embodied in p-n semiconductor junctions. However this simple circuit is not commonly found in isolation in modern circuits. Here, we consider a new version of the RLD circuit, namely a driven RLD circuit followed by an operational amplifier (Trans-Impedance Amplifier - TIA). The RLD-TIA is a more realistic nonlinear circuit element similar to those found in modern electronics. We show in this paper that this combination greatly enhances the susceptibility of the circuit to Radio Frequency Interference (RFI). We show that high power Radio Frequency (RF) signals enhance the nonlinear and chaotic behavior of the circuit at Low Frequency (LF), and this comes from a combination of rectification <u>and</u> nonlinear circuit dynamics. Hence the main focus of this work is to investigate two simple nonlinear driven circuits under RF stimulus and examine the consequences of nonlinear dynamics on circuit behavior.

In Section II we present the experimental setup, describing the circuits tested, variables measured, components and parameters used. Section III presents results on circuit phase diagrams to make contact with the classical literature on the driven RLD circuit [4]-[9],[11]. Section IV-A discusses the observed enhancement of chaotic behavior in the circuit under RF stimulus. There we present bifurcation diagrams and give an explanation of the effect. In Section IV-B we investigate in more detail a DC voltage generation, comparing the voltages measured from the RLD and RLD-TIA circuits. In Section IV-C we present other observed effects from the circuit. At the end, the conclusion section summarizes the main results.

## II. Experiment

There are many ways to investigate chaotic behavior in electronic circuits [13]-[14]. For example, one can examine the state space of variables, such as evolution of the circuit in a two-dimensional space spanned by the electrical current and its time derivative. Also examination of the time series of voltage and current, bifurcation diagram (constructed using local maxima of a state variable), power spectrum, etc. can describe the dynamics associated with the nonlinear system [12]-[14]. In this work we employ bifurcation diagrams constructed from time series measurements of circuit current and voltage. We also measure the power spectrum of the circuit output voltage to map out the system phase diagram.

In addition, when high power microwave signals are involved, care must be taken during the experimental

R. M. de Moraes is with the Center for Superconductivity Research, Department of Physics and also with Department of Electrical and Computer Engineering, University of Maryland, College Park, MD, 20742-4111, USA (e-mail: rmariz@Glue.umd.edu).

S. M. Anlage is with the Center for Superconductivity Research, Department of Physics, University of Maryland, College Park, MD, 20742-4111 USA (e-mail: anlage@squid.umd.edu).

This work was supported by STIC through the STEP Program and by the Department of Defense MURI Program under AFSOR Grant F496 200 110 374. R. M. de Moraes was also supported by CAPES-Brazil.

procedures. The RF and microwave properties of the circuit are very sensitive to parasitic reactances. Hence high impedance and low capacitance probes have been used to reduce parasitic capacitances from being introduced into the circuit. Since the goal of this work is to investigate simple nonlinear low frequency driven circuits under RF stimulus and observe the behavior of the period-doubling and chaotic onset, a controlled impedance environment was created for the circuit. The circuit was assembled on a 50 Ω characteristic impedance circuit board, and all signals were introduced through 50 Ω transmission lines (see Figure 1). All driving signals are CW sinusoidal waveforms. Either the Resistor-Inductor - Diode (RLD circuit, Fig. 1(b)) or the Resistor-Inductor-Diode-Trans-Impedance Amplifier (RLD-TIA circuit, Fig. 1(a)) was the device under test. In the case of the RLD circuit, the variable measured was the voltage drop on the resistor. In the RLD-TIA circuit, the op-amp output voltage ($V_{out}$) was the variable measured. We tested circuits with many different parameters (inductors, resistors, and diodes). Tables I and II show those parameters used and variables measured.

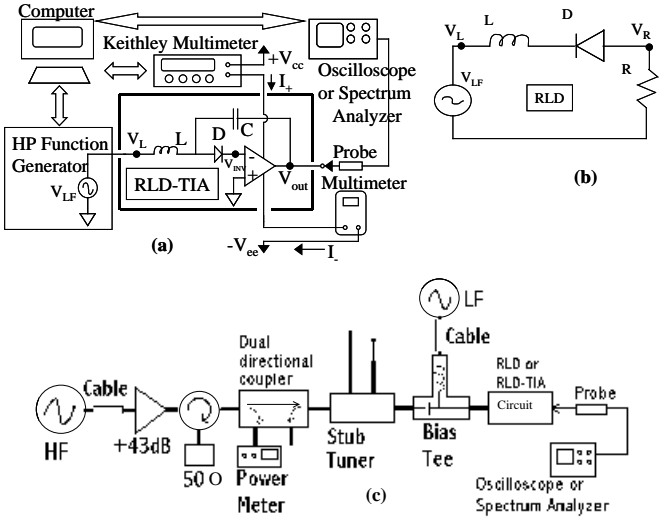

Fig. 1. Experimental setup for high frequency and low frequency study [18]: Low frequency analysis setup for: (a) driven RLD-TIA, (b) Driven RLD circuit, (c) High and low frequency setup together for both circuits.

A LabView program running in the computer controlled the data acquisition process. In the control program we select a LF frequency, an initial LF driving voltage, a final driving voltage, and the step voltage to be swept. At each step, a single shot of the oscilloscope screen presenting the voltage output of the circuit was collected and sent to the computer with time series points stored corresponding to approximately 50 cycles of the fundamental LF drive frequency. After the acquisition, we processed the data generating bifurcation diagrams. The bifurcation diagram was obtained from the time series of the variable measured through a program that searches for local maxima (or minima) and plots them versus the LF voltage amplitude. Figure 1(c) shows the experimental setup used to subject the circuits to both low and high frequency signals. Table II shows typical parameters used in both circuits. The phase diagrams were obtained by observing the power spectrum of the signal from the circuit and determining either the presence of a sub-harmonic of the drive frequency signal in the case of period-doubling, or by the broadband characteristic of the chaotic spectra in the case of chaos.

TABLE I
QUANTITIES MEASURED IN THE EXPERIMENTS.

| Frequency range | Quantities measured in the RLD circuit | Quantities measured in the RLD-TIA circuit |
| --- | --- | --- |
| Low Frequency (1 - 10 MHz), | | $I_+$, $I_-$, $V_{out}$, voltage at inverting input of op-amp $V_{INV}$, DC voltage at inductor $V_L$. See Figure 1(a). |
| Low Frequency (50 – 100 MHz) | Voltage drop on resistor $V_R$, DC voltage at inductor $V_L$. See Figure 1(b). | |
| High Frequency (0.75 - 1 GHz) | $V_R$, $V_L$, See Figure 1(b),1(c). | $I_+$, $I_-$, $V_{out}$, $V_{INV}$, $V_L$, See Figure 1(a),1(c). |

To deliver enough HF power to the circuit we need to impedance match between the source and the circuit. This is accomplished by a stub tuner that is adjusted to maximize the power being delivered to the circuit, and minimize the reflected power. High and low frequency signals are applied to the circuit through a bias tee to prevent the high frequency component from being delivered to the low frequency source generator and *vice-versa*. The high frequencies applied vary from 750 MHz to 1 GHz and power from 0 dBm to +43 dBm. Low frequency signals vary from 1 MHz to 100 MHz and power from -10 dBm to +33 dBm.

TABLE II
PARAMETER VALUES AND COMPONENTS USED IN THE RLD AND RLD-TIA CIRCUITS TESTED.

| Parameters and components | Values, types |
| --- | --- |
| R | 25 Ω, 100 Ω, 1 kΩ, 10 kΩ, 100 kΩ, 1MΩ |
| L | 150 nH, 320 nH, 10 $\mu$H, 100 $\mu$H |
| $|V_{cc}| = |V_{ee}|$ | 6 V, 8 V, 9 V, 12 V, 15 V |
| Diode | NTE610, 1N5475B, forward and reverse biased |
| OP-AMP | MC1741CU, MC1741SCP1 |
| $C_{feedback}$ | 50 pF, 510 pF, 4.7 nF |

III. PHASE DIAGRAM OF THE RLD AND RLD-TIA CIRCUITS

It is helpful to first establish the regions of parameter space where period-doubling and chaos occur. In the circuits of interest to us here, the important parameters (besides the circuit element values) are the driving frequency and amplitude. The phase diagrams presented in Figure 2 indicate the presence of period-doubling or chaos in the LF drive amplitude versus LF drive frequency plane. The "U" shape of this diagram indicates that either period-doubling or chaos are more easily created at frequencies near the resonant frequency of the circuit [11]. The resonant frequency ($f_0$) is estimated by $1/2\pi\sqrt{LC_0}$ where $C_0$ is the transition capacitance of the diode at zero voltage bias [4]-[5],[11]. We observe that for the case of the RLD circuit (L=390 nH, R=25Ω, NTE610 Diode with measured $C_0$=16 pF, yielding $f_0$=70 MHz) only period-doubling is observed, and not chaos, for the frequency range 63 MHz - 83 MHz in Figure 1(b), see Figure 2(a). We also applied high power RF up to +41 dBm to this RLD circuit, but there was no change in the onset of period-doubling for Figure 2(a).

If we add the TIA (Figure 1(a)) with (capacitive) negative feedback ($C_{feedback}$=510 pF, MC1741CU op-amp) to the RLD circuit above, we see that the nonlinear behavior shifts to a much lower frequency range, 1 MHz - 6.5 MHz (see Figure 2(b) - black bars). This is expected because the feedback capacitor acts as if it is in parallel with the varactor diode, giving $C_{Total} \cong C_{feedback}+C_0$, yielding $f_0 \cong 11$ MHz. In this case not only period-doubling was seen but also chaos [19]. From Figure 2 we see that the addition of the TIA feedback enhances the nonlinear effect by bringing down the voltage required to observe period-doubling, and also moves it to much lower frequencies, compared to the RLD circuit above.

by two properly chosen signals may present chaotic instabilities at much lower voltage levels than with one signal alone [15]-[16]. They showed that chaotic behavior can occur when the "high frequency" mode is near the resonant frequency of the circuit and the "low frequency" mode is on the order of the 3dB bandwidth. Our investigation considers a low frequency mode on the order of the resonant frequency of the circuit, while the high frequency mode is in the RF range, about $10^3$ times higher. We are motivated to use this combination by our interest in the combined effects of rectification and nonlinear dynamics, and by the possible generalization of Vavriv's results [15]-[16]. Figure 1(c) shows the experimental setup used.

The general result of adding RF stimulus to the circuit is shown in Figure 2(b)-white bars. The RF causes the system to show period-doubling and chaos at significantly lower LF driving voltage over the entire frequency range. To investigate further, bifurcation diagrams of the driven RLD-TIA circuit are shown in Figure 3(a)-(d) as a function of RF stimulus power. As can be seen in Figure 3(a)-(d), as the incident RF

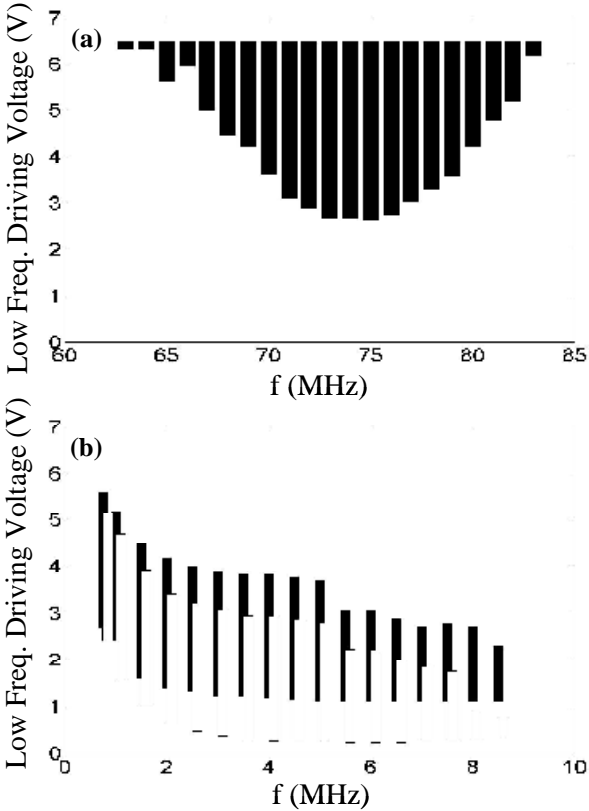

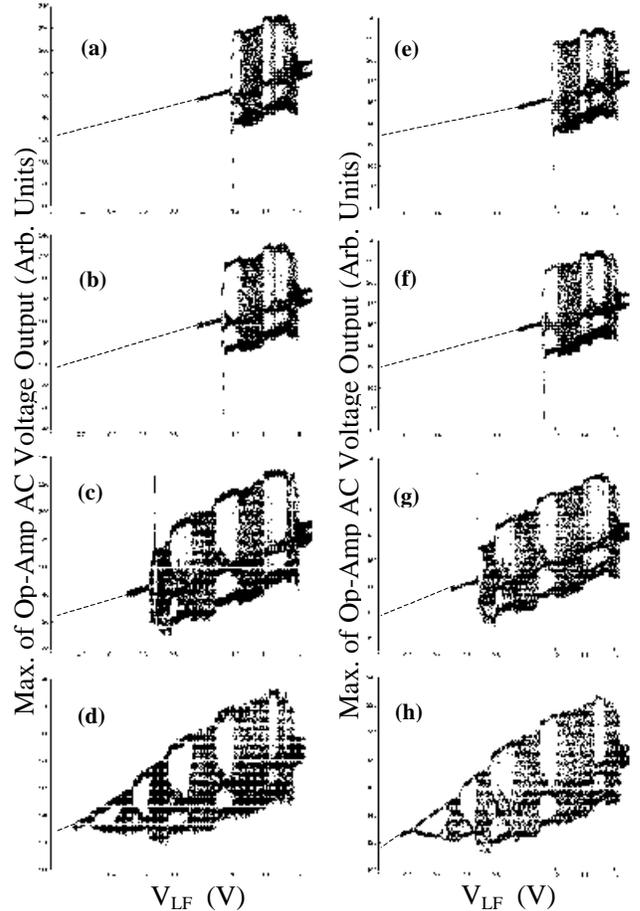

Fig. 2. Period-doubling phase diagrams: (a) driven RLD circuit, where only period-doubling is observed (maximum voltage applied was 6.5 V to prevent damage to the circuit components), (b) RLD-TIA where both chaos and period-doubling are observed (maximum voltage applied was 10 V). Black bars were obtained with no High Frequency signal applied. White bars were obtained with a High Frequency signal applied (f=767 MHz, Power=+33 dBm). The bars represent parameter values where either period-

Fig. 3. Low frequency bifurcation diagrams for RF stimulus (left column) and DC voltage offset (right column) of the RLD-TIA circuit. All graphs show the relative maximum of the op-amp output voltage (using 50 cycles of the driven voltage) plotted versus the amplitude of the low-frequency driving voltage. (a) No High Frequency Power, (b) $P_{HF}$=+20dBm, (c) $P_{HF}$=+30 dBm, (d) $P_{HF}$=+40 dBm RF stimulus. (e) DC offset=0 V, (f) DC Offset=+40 mV, (g) DC offset=+300 mV, (h) DC Offset=+540 mV. (Parameters: L=150 nH, NTE610 Diode forward biased, RF=800 MHz, LF=5.5 MHz, op-amp MC1741CU). Dashed lines indicate where period-1 behavior is observed.

## IV. RF STIMULUS

### A. Enhancement of Chaotic Behavior

It was seen by Vavriv *et al.* that a nonlinear circuit driven power is increased, the nonlinear response at low frequency is enhanced, causing the onset of period-doubling and chaos to begin at lower driving voltages. Note that each bifurcation diagram reproduces all the features of the others at higher LF driving amplitude, to good approximation. This is remarkable because naively we would expect a shift of the bifurcation diagrams by an amount equal to the rectified DC voltage. Instead, the effect of the RF is to "uncover" more and more of the underlying bifurcation diagram. It seems that the RF "tickles" the system out of period 1 behavior into more complicated behavior.

Tanaka *et al.* observed that the driven RLD circuit presents sheet structures in a global bifurcation diagram built using a 3-dimensional state space [17]. The first two state variables are those shown in the phase diagram of Figure 2, and the third variable used there is S/T, where S represents the time the trajectory in the state space spends in the diffusion capacitance region of the diode ($C_d$), while T stands for the time it spends in the transition capacitance ($C_t$) region [17] (see Figure 4). Our results resemble theirs, where the third dimension in our case can be explored through the high frequency power. Hence the "uncovering" feature observed in Figure 3(a)-(d) may be a consequence of jumping over the bifurcation sheet structures as we change the RF power.

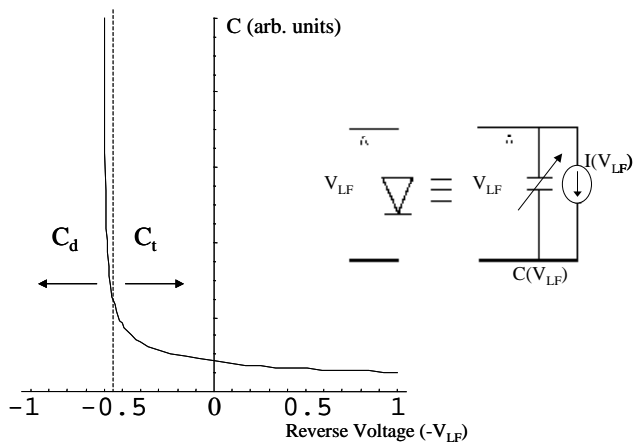

Fig. 4. General schematic diode circuit model. The crossover between transition capacitance $C_t$ and diffusion capacitance $C_d$ occurs when the charge on the "capacitor plate" goes through zero [5],[17]. The forward DC offset voltage brings the circuit operation to a region close to where $C(V_{LF})$ is very nonlinear. $I(V_{LF})$ is the nonlinear diode current-voltage relationship [4]-[5].

In order to understand the RF power dependence of the RLD-TIA circuit, we replaced the RF signal with a DC offset voltage. The resulting bifurcation diagrams are presented in Figure3(e)-(h) for several values of applied DC offset voltage. If we compare those results to Figure 3(a)-(d) we see nearly identical behavior, suggesting that the RF signal is equivalent to a DC offset voltage applied to the diode and op-amp. This can be understood as follows. As the forward biased DC offset voltage across the diode is increased we are getting closer to the very nonlinear region of the model diode curve $C(V_{LF})$ (see Figure 4) [20]. Hence only relatively small amplitudes of low frequency drive voltage are required to reach that region. We tested this hypothesis by applying a negative bias to the diode and we observed the opposite behavior, i.e. a "recovering" of the bifurcation diagram forcing us to apply a larger LF driving voltage to see period-doubling and chaos in the circuit.

We repeated this experiment for the RLD circuit using parameters and conditions where LF chaotic behavior is present (e.g. L=10 μH, R=25 Ω, and NTE610 diode), and we did not observe significant changes in the onset of chaos due to the applied RF.

### B. Measurement of Rectified DC Offset Voltage

In the previous section we observed that the nonlinear properties of the RLD-TIA circuit can be modified if we apply a tone with a frequency much greater than the natural resonant frequency of the circuit. More specifically, the RF component seems to be equivalent to a DC source voltage when the operational amplifier adds a nonlinear feedback in the RLD circuit. Therefore we did an experiment to measure the DC voltage level obtained as a result of the RF stimulus. Figure 5 shows the DC voltage offset measured at the inductor ($V_L$ in Figures 1(a),1(b)) using the setup of Figure 1(c) where the circuit inside the box was either the RLD or the RLD-TIA. The low frequency source generator was replaced by a digital multimeter to measure the DC voltage level through the bias tee. We observe that the voltage levels developed in the RLD-TIA are much higher than in the case of the RLD circuit for the same HF power (even using other values of resistance, see Table II). Clearly the DC voltage is enhanced by the presence of the op-amp and feedback. These results, together with the observations from the previous section, suggest that the negative feedback and the RF stimulus cause the circuit to develop a DC voltage level at the inductor, thus bringing the circuit to a region where the diode capacitance is more nonlinear (see Figure 4), resulting in the onset of chaotic behavior at lower values of LF drive voltage.

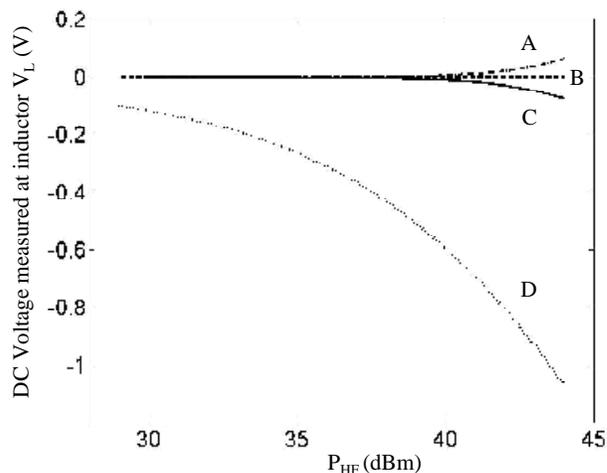

Fig. 5. DC Offset Voltage measured at the inductor $V_L$ versus RF stimulus power: L=390 nH, NTE610 Diode, HF=800 MHz. RLD circuit: R=25 Ω . RLD-TIA circuit: op-amp MC1741CU. A) Diode reverse biased - RLD, B) Diode reverse biased - RLD-TIA, C) Diode forward biased - RLD, D) Diode forward biased - RLD-TIA.

### C. Other Observations

In addition to the observations made above, unusual behavior at period doubling and chaotic transitions of the RLD-TIA circuit were found. We observed discontinuities in the dc supply current to the op-amp at each period doubling transition and chaotic transition. Van Buskirk *et al.* observed that the bifurcation diagram of the average current in the driven RLD circuit also presents discontinuities at transitions between periodic and chaotic behavior [7]. However, the calculated dissipated power associated with the supply current discontinuities is far too small to create irreversible changes in the op-amp by thermal means. In no case have we ever observed failure or other irreversible changes of the op-amp under any experimental circumstances considered thus far.

PSpice simulations were done for the RLD circuit. We have qualitative agreement, for example, with Figure 2(a), although the simulation shows period-doubling for frequencies below 60 MHz and above 85 MHz. Nevertheless the "U" shape is clearly observed, and the voltage scales are similar.

Also the positive DC level branch measured for a reverse-biased diode in the RLD circuit (curve A in Figure 5) is not observed for the reverse-biased diode with the operational amplifier present (RLD-TIA circuit) (curve B in Figure 5). Hence diode bias symmetry is not preserved when an operational amplifier with negative feedback is present, and this explains why we need to have the diode forward biased in the circuit (see Figure 1(a)) to observe nonlinear behavior.

Although most of the results presented here were obtained using the NTE610 diode and 150 nH and 320 nH inductors in the RLD-TIA circuit, we also observed similar results for the varactor diode 1N5475B, suggesting that the enhancement in the nonlinear behavior is a general property.

## V. CONCLUSIONS

We have examined the canonical RLD circuit modeling the nonlinear dynamics of a p-n junction, along with a trans-impedance amplifier (TIA), and found a striking change in the onset of period-doubling under RF stimulus. Phase diagrams were presented showing that chaos and bifurcations were obtained when the TIA was added to the circuit. High RF power enhances the nonlinear and chaotic behavior at lower LF voltages for the RLD-TIA circuit. This is associated with a DC voltage induced by rectification in the circuit, which was verified in both the RLD circuit and RLD-TIA circuits. The combination of rectification and nonlinear characteristics can thus enhance the range of chaotic behavior of the circuit. Rectification is only part of the effect of an RF stimulus signal on a nonlinear circuit. The nonlinear dynamics of the circuit can greatly enhance the disruptive effect of the RF stimulus. This suggests that qualitatively new behavior may be found when p-n junctions are embedded in modern electronics.

## ACKNOWLEDGMENT

We acknowledge helpful discussions with Thomas Carroll, and William Crevier. Early work on the circuit was done by Sang-Ho Bok.